\documentclass{ws-procs10x7}
\usepackage{balance}

\newcolumntype{d}[1]{D{.}{.}{#1}}

\makeindex
\begin{document}

\title{Can we learn something more on oscillations from atmospheric neutrinos?}

\author{T. SCHWETZ}

\address{Scuola Internazionale Superiore di Studi Avanzati,
         Via Beirut 2--4, I--34014 Trieste, Italy}

\twocolumn[\maketitle\abstract{We show that for long-baseline
experiments using a Mt water \v{C}erenkov detector atmospheric neutrino
data provide a powerful method to resolve parameter degeneracies. In
particular, the combination of long-baseline and atmospheric data
increases significantly the sensitivity to the neutrino mass hierarchy
and the octant of $\theta_{23}$. Furthermore, we discuss the
possibility to use $\mu$-like atmospheric neutrino data from a big
magnetized iron calorimeter to determine the neutrino mass hierarchy.}]

\section{Introduction}

In the past atmospheric neutrinos have played an important role in
establishing the phenomenon of neutrino oscillations\cite{SKatm}.
However, now we are entering the era of high precision long-baseline
(LBL) experiments\cite{lbl,t2k,Huber:2004ug,c2m} which will outperform
atmospheric neutrinos in the determination of the oscillation
parameters $|\Delta m^2_{31}|$ and $\sin^22\theta_{23}$. Hence, the
question raised in the title of this talk arises. In the following I
will suggest that the answer to this question is ``yes'', by
discussing possibilities to use atmospheric neutrino data from Mt
scale water \v{C}erenkov detectors\cite{Wcerenkov} or from large
magnetized iron calorimeters\cite{INO} (see also
Ref.\cite{Choubey:2006jk} for a recent review).

\section{Combining LBL and ATM data from Mt water detectors}

The primary aims of future neutrino experiments are the determination
of the mixing angle $\theta_{13}$, the CP-phase $\delta_\mathrm{CP}$,
and the type of the neutrino mass hierarchy (normal or inverted),
i.e., the sign of $\Delta m^2_{31}$. It is well known that parameter
degeneracies are a severe problem on the way towards these goals. In
Ref.\cite{Huber:2005ep} it was demonstrated that for LBL experiments
based on Mt scale water \v{C}erenkov detectors data from atmospheric
neutrinos (ATM) provide an attractive method to resolve degeneracies.

Atmospheric neutrinos are sensitive to the neutrino mass hierarchy if
$\theta_{13}$ is sufficiently large due to Earth matter effects,
mainly in multi-GeV $e$-like events\cite{atm13}. Moreover, sub-GeV
$e$-like events provide sensitivity to the octant of
$\theta_{23}$~\cite{Peres:2003wd,Gonzalez-Garcia:2004cu} due to
oscillations with $\Delta m^2_{21}$. However, these effects can be
explored efficiently only if LBL data provide a very precise
determination of $|\Delta m^2_{31}|$ and $\sin^22\theta_{23}$, as well
as some information on $\theta_{13}$~\cite{Huber:2005ep}.

\begin{figure*}[!ht]
\centering
\includegraphics[width=0.88\textwidth]{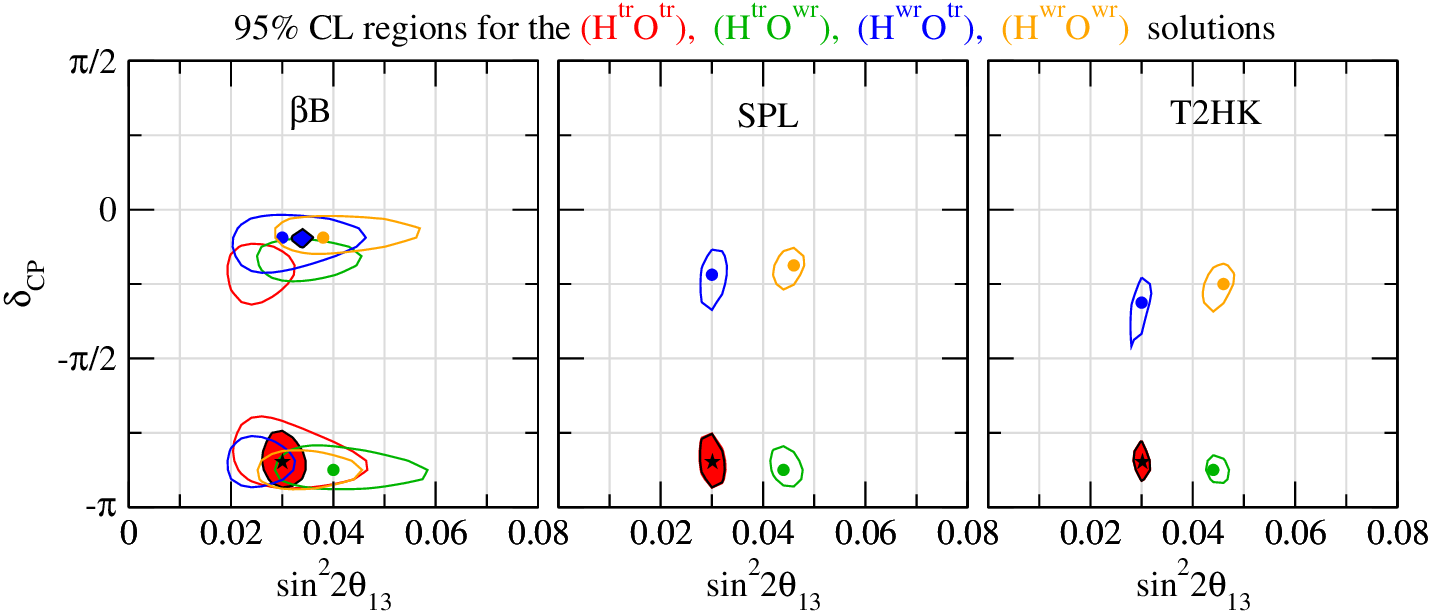}
\caption{Allowed regions in $\sin^22\theta_{13}$ and
  $\delta_\mathrm{CP}$ for LBL data alone (contour lines) and LBL+ATM
  data combined (colored regions). The true parameter values are
  $\delta_\mathrm{CP} = -0.85 \pi$, $\sin^22\theta_{13} = 0.03$,
  $\sin^2\theta_{23} = 0.6$. $\mathrm{H^{tr/wr} (O^{tr/wr})}$
  refers to solutions with the true/wrong mass hierarchy (octant of
  $\theta_{23}$).}
\label{fig:degeneracies}
\end{figure*}
 
Here we illustrate the synergies from a combined LBL+ATM analysis at
the examples of the T2K phase II experiment\cite{t2k} (T2HK) with the
HK detector of 450~kt fiducial mass, and two experiments with beams
from CERN to a 450~kt detector at Frejus (MEMPHYS)\cite{c2m}, namely
the SPL super beam and a $\gamma = 100$ beta beam ($\beta$B). The LBL
experiments are simulated with the GLoBES software\cite{globes}, and a
general three-flavor analysis of ATM data is
performed\cite{c2m,Huber:2005ep,Gonzalez-Garcia:2004cu}. For each
experiment we assume a running time of 10 years, where the
neutrino/anti-neutrino time is chosen as 2+8 years for SPL and T2K, and
5+5 years for the beta beam, see Ref.\cite{c2m} for details.

The effect of degeneracies becomes apparent in
Fig.~\ref{fig:degeneracies}. For given true parameter values the data
can be fitted with the wrong hierarchy and/or with the wrong octant of
$\theta_{23}$. Hence, from LBL data alone the hierarchy and the octant
cannot be determined. Moreover, as visible from the solid lines in
Fig.~\ref{fig:degeneracies} the degenerate solutions appear at
parameter values different from the true ones, an hence, ambiguities
exist in the determination of $\theta_{13}$ and $\delta_\mathrm{CP}$.
If the LBL data are combined with ATM data only the colored regions in
Fig.~\ref{fig:degeneracies} survive, i.e., in this particular example
for all three experiments the degeneracies are completely lifted at
95\%~CL, the mass hierarchy and the octant of $\theta_{23}$ can be
identified, and the ambiguities in $\theta_{13}$ and
$\delta_\mathrm{CP}$ are resolved. Let us note that here we have
chosen a favorable value of $\sin^2\theta_{23} = 0.6$; for values
$\sin^2\theta_{23} < 0.5$ in general the sensitivity of ATM data is
weaker\cite{Huber:2005ep}.

\begin{figure}[!ht]
\centering
\includegraphics[width=0.44\textwidth]{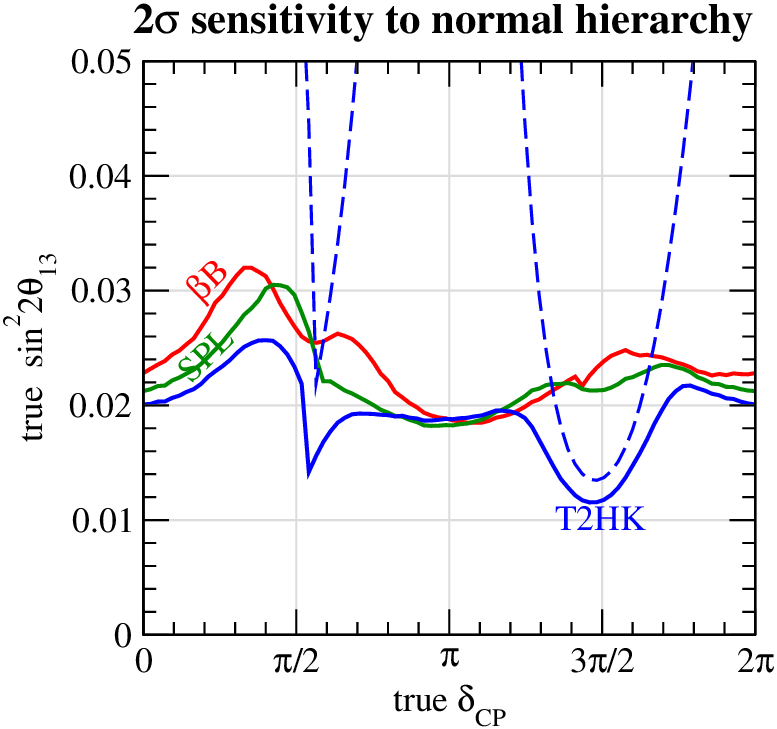}
\caption{Sensitivity to the neutrino mass hierarchy as a function of
   $\sin^22\theta_{13}$ and $\delta_\mathrm{CP}$ for
   $\theta_{23}^\mathrm{true} = \pi/4$ and a true normal
   hierarchy. Solid curves correspond to LBL+ATM data combined, the
   dashed curves correspond to T2HK LBL data-only. $\beta$B and SPL
   without ATM have no sensitivity to the hierarchy.}
\label{fig:hierarchy}
\end{figure}

In Fig.~\ref{fig:hierarchy} we show the sensitivity to the neutrino
mass hierarchy. For LBL data alone there is practically no sensitivity
for the CERN--MEMPHYS experiments (because of the very small matter
effects due to the relatively short baseline of 130~km), and the
sensitivity of T2HK depends strongly on the true value of
$\delta_\mathrm{CP}$. However, with the LBL+ATM combination all
experiments can identify the mass hierarchy at $2\sigma$~CL provided
$\sin^22\theta_{13} \gtrsim 0.02-0.03$~\cite{c2m}.

\begin{figure}[!t]
\centering
  \includegraphics[width=0.44\textwidth]{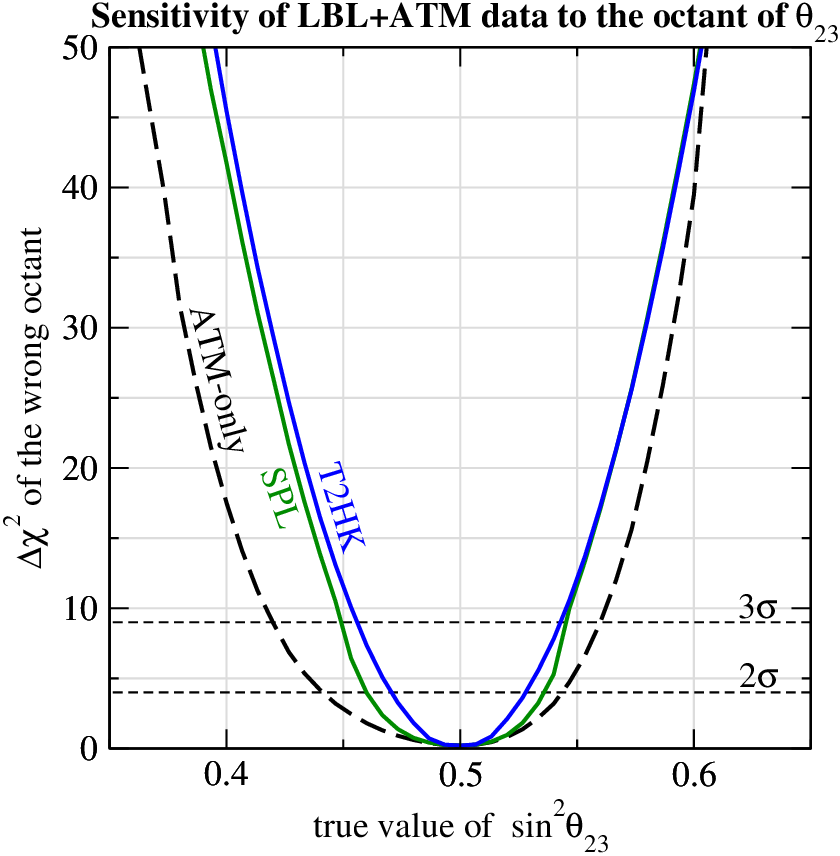}
  \caption{$\Delta\chi^2$ of the solution with the wrong octant of
  $\theta_{23}$ as a function of $\sin^2\theta_{23}$. We have assumed
  a true value of $\theta_{13} = 0$.}
  \label{fig:octant}
\end{figure}

Fig.~\ref{fig:octant} shows the potential of ATM+LBL data to exclude
the octant degenerate solution. Since this effect is based mainly on
oscillations with $\Delta m^2_{21}$ there is very good sensitivity
even for $\theta_{13} = 0$; a finite value of $\theta_{13}$ in general
improves the sensitivity\cite{Huber:2005ep}.  From the figure one can
read off that atmospheric data alone can resolve the correct
octant at $3\sigma$ if $|\sin^2\theta_{23} - 0.5| \gtrsim 0.085$. If
atmospheric data is combined with the LBL data from SPL or T2HK there
is sensitivity to the octant for $|\sin^2\theta_{23} - 0.5| \gtrsim
0.05$.

\section{Magnetized iron calorimeters}

In water \v{C}erenkov detectors one cannot distinguish between
neutrino and anti-neutrino events. This limits the sensitivity to the
mass hierarchy, since depending on the hierarchy the resonance occurs
either for neutrinos or anti-neutrinos. Therefore, in principle one
expects that the sensitivity improves for detectors capable to
distinguish atmospheric neutrino from anti-neutrino events. In the
following we discuss the possibility offered by a large (several
10~kt) magnetized iron calorimeter similar to the INO
proposal\cite{INO}. Such a detector can determine the charge of muons,
whereas electron detection is difficult. The principles of atmospheric
neutrino measurements with a 5.4~kt detector of this type have been
established recently by the MINOS experiment\cite{Adamson:2005qc}.

Here we report the results obtained in Ref.\cite{Petcov:2005rv}, see
Ref.\cite{magn} for related considerations.  We limit ourselves to
$\mu$-like events, and we assume a correct identification of
$\nu_\mu$- versus $\bar\nu_\mu$-events of 95\%. The observation of the
muon and the hadronic event allows in principle to reconstruct the
original direction and energy of the neutrino. Indeed, it has been
stressed in Ref.\cite{Petcov:2005rv} that the accuracy of neutrino
energy and direction reconstruction is crucial for the determination
of the hierarchy. The reason is that the difference in the event
spectra of normal and inverted hierarchy show a characteristic
oscillatory pattern. If this pattern can be resolved a powerful
discrimination between the hierarchies is possible. If however, the
oscillatory pattern is washed out because of a poor accuracy in energy
and direction reconstruction the sensitivity to the hierarchy
decreases drastically.

\begin{figure*}[!t]
\centering
  \includegraphics[width=0.75\textwidth]{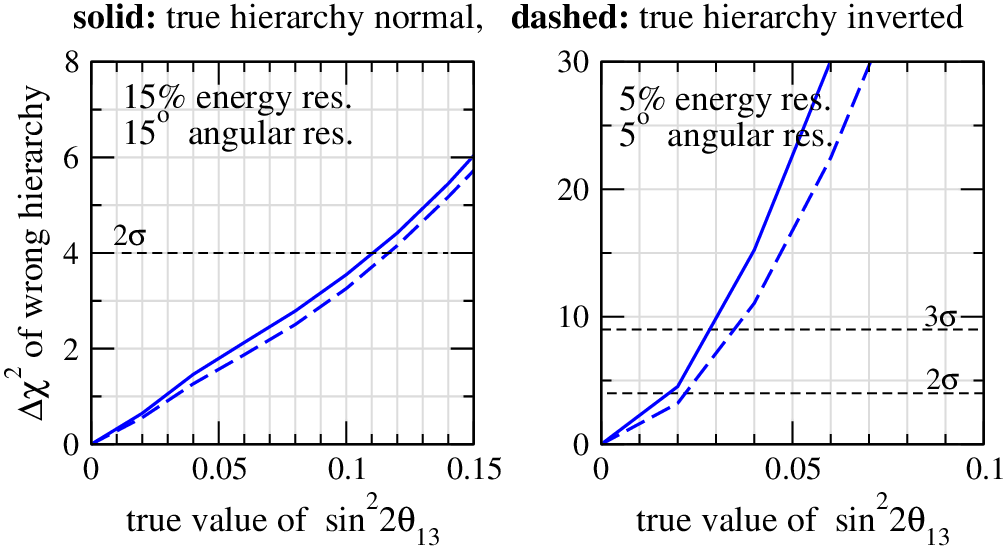}
  \caption{Sensitivity to the neutrino mass hierarchy of a magnetized
  iron calorimeter for different assumptions on the neutrino energy
  and direction reconstruction accuracy. We assume 500 kt yr data from
  an INO-like detector, corresponding to 4000 up-going $\mu$-like
  events with $E_\mu > 2$~GeV. Here, $\theta_{23}^\mathrm{true} =
  \pi/4$ and we assume external information on $|\Delta m^2_{31}|$ and
  $\sin^22\theta_{23}$ of 10\%, and an uncertainty on
  $\sin^22\theta_{13}$ of $\pm 0.02$.}
  \label{fig:magn}
\end{figure*}

In Fig.~\ref{fig:magn} we show the sensitivity to the hierarchy for a
500~kt~yr exposure of an INO-like detector. In the left panel we
assume that the neutrino energy can be reconstructed with an accuracy
of 15\% and the neutrino direction with an accuracy of $15^\circ$,
whereas in the right panel the very optimistic accuracies of 5\% and
$5^\circ$ are adopted. Details on our simulation and systematic errors
are given in Ref.\cite{Petcov:2005rv}. One observes from the plot that
for optimistic assumptions the hierarchy can be identified at
$2\sigma$ if $\sin^22\theta_{13} \gtrsim 0.02$. This sensitivity is
comparable to the one from Mt water \v{C}erenkov detectors discussed in
the previous section. If however, more realistic values for the energy
and direction reconstruction are adopted the sensitivity deteriorates
drastically and values of $\sin^22\theta_{13} \gtrsim 0.1$ (close to
the present bound) are required.

\section{Conclusions}

In this talk I have discussed the potential of future atmospheric
neutrino experiments. An interesting possibility arises if a LBL
super beam or beta beam experiment with a Mt scale water
\v{C}erenkov detector is built. In such a case the high statistics atmospheric
neutrino data available in the detector may provide complementary
information to the LBL data and help to resolve parameter
degeneracies. In particular, the sensitivity to the neutrino mass
hierarchy and the octant of $\theta_{23}$ is significantly increased.
Furthermore, I have discussed large magnetized iron calorimeter
detectors. Such detectors are under consideration as far detector for
a neutrino factory, and can be viewed as an up-scaled version of the
present MINOS experiment. For the determination of the neutrino mass
hierarchy with $\mu$-like events in a magnetized iron detector the
ability to reconstruct the neutrino energy and direction with good
accuracy is crucial.

{\bf Acknowledgment.} T.S.\ is supported by the $6^\mathrm{th}$~Framework
Program of the European Community under a Marie Curie Intra-European
Fellowship.


\begin{thebibliography}{99}

\bibitem{SKatm}
  Super-K Collaboration,
  Y.~Fukuda {\it et al.}, 
  Phys.\ Rev.\ Lett.\  {\bf 81} (1998) 1562; 
%
  Y.~Ashie {\it et al.},
  Phys.\ Rev.\ D {\bf 71} (2005) 112005;
%
  C.~Walter, these proceedings.

\bibitem{lbl}
  D.G.Micheal {\it et al.} [MINOS],
  hep-ex/0607088,
%
  G.~Pearce, these proceedings;
%
  D.S.~Ayres {\it et al.} [NOvA],
  hep-ex/0503053,
%
  L.~Mualem, these proceedings.

\bibitem{t2k}
  Y.~Itow {\it et al.},
  hep-ex/0106019,
%
  Yu.~Kudenko, these proceedings.

\bibitem{Huber:2004ug}
  P.~Huber {\it et al.}, 
  Phys.\ Rev.\ D {\bf 70} (2004) 073014
  [hep-ph/0403068].

\bibitem{c2m}
  J.~E.~Campagne, M.~Maltoni, M.~Mezzetto and T.~Schwetz,
  hep-ph/0603172.

\bibitem{Wcerenkov}
  C.~K.~Jung,
  hep-ex/0005046;
%
  K.~Nakamura,
  Int.\ J.\ Mod.\ Phys.\ A {\bf 18} (2003) 4053;
%
  A.~de Bellefon {\it et al.}, 
  hep-ex/0607026.

\bibitem{INO} G.~Rajasekaran,
  AIP Conf.\ Proc.\  {\bf 721} (2004) 243;
%
  A.~Dighe, these proceedings.

\bibitem{Choubey:2006jk}
  S.~Choubey,
  hep-ph/0609182.

\bibitem{Huber:2005ep}
  P.~Huber, M.~Maltoni, T.~Schwetz,
  Phys.\ Rev.\ D {\bf 71} (2005) 053006
  [hep-ph/0501037].

\bibitem{atm13}
  E.K.~Akhmedov {\it et al.}, 
  Nucl.\ Phys.\ B {\bf 542} (1999) 3;
%
  J.~Bernabeu, S.~Palomares Ruiz, S.T.~Petcov,
  Nucl.\ Phys.\ B {\bf 669} (2003) 255.

\bibitem{Peres:2003wd}
  O.L.G.~Peres, A.Y.~Smirnov,
  Nucl.\ Phys.\ B {\bf 680} (2004) 479
  [hep-ph/0309312].

\bibitem{Gonzalez-Garcia:2004cu}
  M.C.~Gonzalez-Garcia, M.~Maltoni, A.Y. Smirnov,
  Phys.\ Rev.\ D {\bf 70} (2004) 093005.

\bibitem{globes}
  P.~Huber, M.~Lindner, W.~Winter,
  Comput.\ Phys.\ Commun.\  {\bf 167} (2005) 195
  [hep-ph/0407333],
  www.ph.tum.de/$\sim$globes

\bibitem{Adamson:2005qc}
  P.~Adamson {\it et al.}  [MINOS Coll.],
  Phys.\ Rev.\ D {\bf 73}, 072002 (2006)
  [hep-ex/0512036].

\bibitem{Petcov:2005rv}
  S.~T.~Petcov and T.~Schwetz,
  Nucl.\ Phys.\ B {\bf 740}, 1 (2006)
  [hep-ph/0511277].

\bibitem{magn}
  T.~Tabarelli de Fatis,
  Eur.\ Phys.\ J.\ C {\bf 24}, 43 (2002);
%
  S.~Palomares-Ruiz and S.~T.~Petcov,
  Nucl.\ Phys.\ B {\bf 712} (2005) 392;
%
  D.~Indumathi and M.~V.~N.~Murthy,
  Phys.\ Rev.\ D {\bf 71} (2005) 013001;
%
  R.~Gandhi {\it et al.}, 
  Phys.\ Rev.\ D {\bf 73} (2006) 053001;
%
  S.~Choubey and P.~Roy,
  Phys.\ Rev.\ D {\bf 73}, 013006 (2006).



\end{thebibliography}
\end{document}